\documentclass[%
 reprint,
 amsmath,amssymb,
 aps,
 superscriptaddress,
 prl
]{revtex4-1}

\usepackage{graphicx}% Include figure files
\usepackage{dcolumn}% Align table columns on decimal point
\usepackage{bm}% bold math
\usepackage{xcolor}
\usepackage{siunitx}
\usepackage{amsmath}
\usepackage{float}

\begin{document}
%TC:ignore
\preprint{APS/123-QED}

\title{Multipartite entanglement in a microwave frequency comb}

\author{Shan W. Jolin}
\email{shan@meetiqm.com}
\altaffiliation{Present address: IQM Finland Oy, Espoo 02150, Finland}
\affiliation{Department of Applied Physics, KTH Royal Institute of Technology, SE-106 91 Stockholm, Sweden}

\author{Gustav Andersson}
\affiliation{Pritzker School of Molecular Engineering, University of Chicago, Chicago, Illinois 60637, USA}
\affiliation{Department of Microtechnology and Nanoscience MC2, Chalmers University of Technology, SE-412 96 G{\"o}teborg, Sweden}

\author{J. C. Rivera Hern{\'a}ndez}
\affiliation{Department of Applied Physics, KTH Royal Institute of Technology, SE-106 91 Stockholm, Sweden}

\author{Ingrid Strandberg}
\affiliation{Department of Microtechnology and Nanoscience MC2, Chalmers University of Technology, SE-412 96 G{\"o}teborg, Sweden}

\author{Fernando Quijandr{\'i}a}
\altaffiliation{Present address: Quantum Machines Unit, Okinawa Institute of Science and Technology Graduate University, Onna-son, Okinawa 904-0495, Japan}
\affiliation{Department of Microtechnology and Nanoscience MC2, Chalmers University of Technology, SE-412 96 G{\"o}teborg, Sweden}

\author{Jos\'{e} Aumentado}
\affiliation{National Institute of Standards and Technology, 325 Broadway, Boulder, CO 80305, USA}

\author{Riccardo Borgani}
\affiliation{Department of Applied Physics, KTH Royal Institute of Technology, SE-106 91 Stockholm, Sweden}
\affiliation{Intermodulation Products AB, SE-823 93 Segersta, Sweden}

\author{Mats O. Thol{\'e}n}
\affiliation{Department of Applied Physics, KTH Royal Institute of Technology, SE-106 91 Stockholm, Sweden}
\affiliation{Intermodulation Products AB, SE-823 93 Segersta, Sweden}

\author{David B. Haviland}
\affiliation{Department of Applied Physics, KTH Royal Institute of Technology, SE-106 91 Stockholm, Sweden}
\email{haviland@kth.se}

\date{\today}

\begin{abstract}
Significant progress has been made with multipartite entanglement of discrete qubits, but continuous variable systems may provide a more scalable path toward entanglement of large ensembles. We demonstrate multipartite entanglement in a microwave frequency comb generated by a Josephson parametric amplifier subject to a bichromatic pump. We find 64 correlated modes in the transmission line using a multifrequency digital signal processing platform. Full inseparability is verified in a subset of seven modes. Our method can be expanded to generate even more entangled modes in the near future.
\end{abstract}

\maketitle

%TC:endignore

Superconducting quantum circuits have demonstrated excellent unitary control and coherence of qubits at a level sufficient to usher in the era of noisy intermediate-scale quantum technologies \cite{Arute2019}.
So far, a vast majority of the research is focused on the circuit-based approach to quantum computing \cite{Nielsen2000}.
An alternative paradigm is the measurement-based one-way quantum computer \cite{Raussendorf2001, Raussendorf2003, Larsen2020}, where computation is realized through a sequence of measurements on so-called cluster states \cite{Briegel2001, Menicucci2006}.
This paradigm has received significantly more attention in optics, where the continuous variable (CV) counterpart has revealed large scale multipartite entanglement in optical frequency combs involving many thousands of modes \cite{Roslund2014, Pfister2019, Yokoyama2013, Yoshikawa2016, Larsen2019, Asavanant2019}.

Ever since the pioneering experiments of Yurke et. al. \cite{Yurke1988, Movshovich1990}, single and two-mode squeezing has been observed in a wide variety of systems,  from superconducting \cite{tholen2009, Eichler2011, Beltran2008, Flurin2012, Eichler2014, Wilson2011, Menzel2012, Lahtenmaaki2013, Schneider2020, Perelshtein2021, Esposito2022, Qiu2022} to mechanical \cite{Palomaki2013, Ockeloen2017, Kotler2021}.
However, to generate an arbitrary cluster state, multimodal squeezing \cite{Zipilli2020} beyond two modes is required.
Tripartite squeezing has been demonstrated in superconducting devices \cite{Chang2018, Agusti2020, Laehteenmaeki2016} and multimode squeezing has been demonstrated with parametrically coupled surface acoustic wave modes \cite{andersson2021}.
But large multipartite entanglement in the microwave spectrum with superconducting circuits remains elusive.

In this work we demonstrate squeezing of multiple propagating modes in a transmission line connected to a Josephson parametric amplifier (JPA) with a bichromatic pump.
Our digital signal processing platform enables measurements of correlations between as many as 64 modes. While genuine multipartite entanglement between so many modes is nontrivial to establish unequivocally \cite{Sperling2013, Shchukin2015, Gerke2015, Gerke2016}, we present compelling evidence of seven fully inseparable modes. Our method provides a clear path for scaling to many more modes and perhaps construction of CV cluster states.

\begin{figure}
    \centering
     \includegraphics[width=\columnwidth]{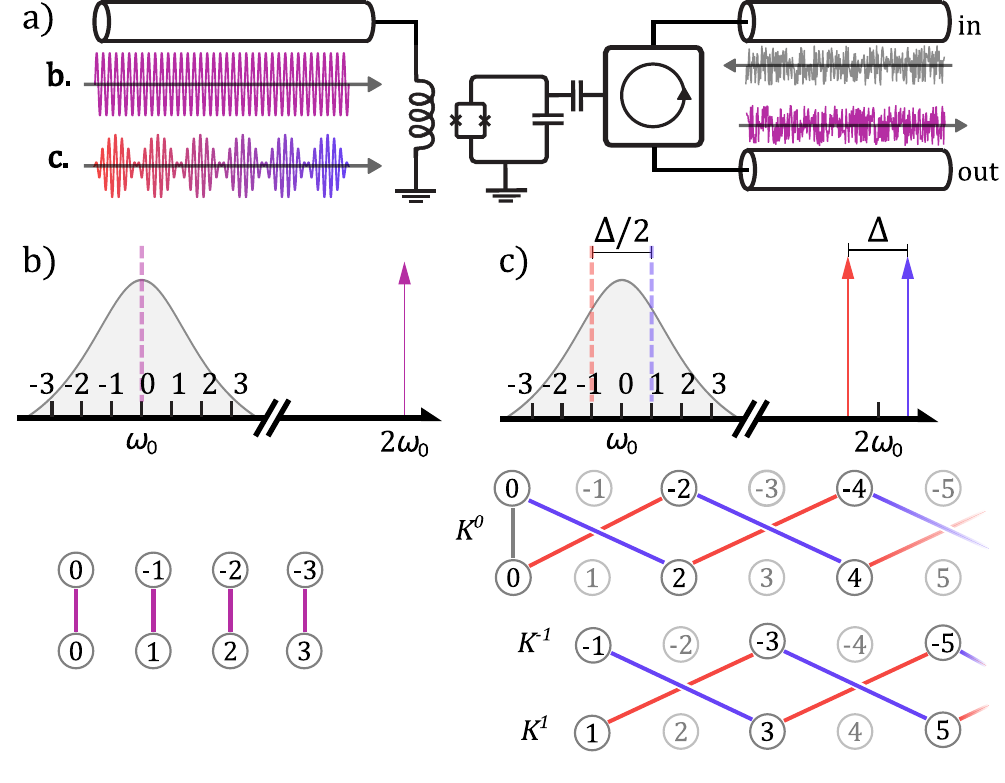}
    \caption{a) The JPA forms a LC circuit with tunable inductance $L_J$ due to the SQUID. The inductance, and hence its resonance frequency $\omega_0$, is modulated by an external flux field of the monochromatic or bichromatic kind, labelled \textbf{b} and \textbf{c} respectively. The pump stimulates incoming vacuum noise scattering off the JPA resonance to become entangled (right). The incoming and outgoing noise is separated by a circulator. \newline
    b) A monochromatic flux pump is applied at $\omega_p \approx 2 \omega_0$, which correlates modes symmetrically around mode 0.
    We use a graph to indicate the resulting correlations in the frequency comb. Numbered vertices correspond to labels on the frequency axis and edges indicate classical or quantum correlations.
    \newline
    c) A bichromatic flux pump consists of two different frequencies at roughly $2\omega_0$.
    We can illustrate the correlations with graphs $K^0$, $K^{-1}$ and $K^1$. The superscript indicates the root vertex.
    In our paper, $\Delta = 2\pi \cdot \SI{9.2}{MHz}$ and $\omega_0 \approx 2\pi \cdot \SI{4.3}{GHz} $.}
    \label{fig:physics_schematic}
\end{figure}

Our JPA is a lumped-element LC circuit (see Fig.~\ref{fig:physics_schematic}) cooled to \SI{10}{mK}, where the inductor is replaced by a superconducting quantum interference device (SQUID). 
The JPA is over-coupled at the signal port to a circulator which separates incoming and outgoing modes in two transmission lines, with the outgoing modes connected through a double isolator to a cryogenic low-noise amplifier. 
The second port is inductively coupled to the SQUID loop, through which we apply a time-varying flux pump to modulate the circuit inductance.
The flux pump together with the flux bias permit a 3-wave mixing process, known to amplify small signals and generate two-mode squeezing \cite{Roy2016}.
The over-coupled linewidth of the JPA resonance is $\kappa_\text{total} = 2\pi \cdot \SI{124}{MHz}$.

Noise quadrature data are collected by a digital multi-frequency lockin that simultaneously demodulates at many frequencies, each being an integer multiple of the measurement bandwidth.
We measure the \textit{IQ}-quadratures of up to 64 evenly-spaced frequencies with no Fourier leakage between the demodulated frequencies.
The measurement is facilitated by directly digitizing in the second Nyquist zone, granting access to the \SI{2.5}{} -- \SI{5}{GHz} band without analog mixers. 
Analog bandpass filters at the lockin input reduce the effect of aliasing.
The digital multi-frequency lockin's clock is the reference oscillation for our external microwave generator which supplies the flux pump.
For additional details, see the Supplementary materials, which includes Refs.~\cite{intermodulation_products, tholenRSI_2011, imp_manual, Jolin2020, Mariantoni2010, Tholen_thesis, Clerk2010, Caves1982, Pastore2019, diamond2016, agrawal2018, diff_evolution, scipy_kurtosis, scipy_skewness, Gardiner1985, Hong2016Apr, Huber2010May, Shalm2013Jan, Teh2014Dec, Guhne2009Apr, Aoki2003Aug, Serafini2017Oct, Hyllus2006Apr}.

The frequencies at which we measure and demodulate, the modes, are determined by the flux pump frequencies.
A bichromatic pump, with frequencies $\Omega_1$ and $\Omega_2$, is applied at the flux bias port.
The frequencies are approximately centered at twice the JPA resonance $\tfrac{1}{2}(\Omega_2 + \Omega_1) \approx 2 \omega_0$ with a detuning $\Delta$ smaller than the JPA linewidth, $|\Omega_2 - \Omega_1| = \Delta \ll \kappa_{\text{total}}$. 
Measurement is carried out at a comb of frequencies $f_n = \tfrac{1}{4}(\Omega_2 + \Omega_1) + n \Delta/4$, $n \in \mathcal{Z}$, with spacing $\Delta/4$ (see Fig.~\ref{fig:physics_schematic}c). 
This particular frequency selection ensures that signal-idler pairs associated with each pump overlap, giving rise to multimode squeezing.
In our experiments we use $\Delta = 2\pi \cdot \SI{9.2}{MHz}$ for the pumps and $\Delta/4 = 2\pi \cdot \SI{2.3}{MHz}$ for the frequency comb.

To help visualize the correlations induced by pumping, we first consider the familiar case of a monochromatic pump, as illustrated in Fig.~\ref{fig:physics_schematic}b.
Every frequency in the comb is labelled by an integer, with 0 signifying the center frequency $(\Omega_1 + \Omega_2)/4$.
In the 3-wave mixing process  $2 \omega_0 = \omega_n + \omega_{-n}$, correlations arise between signal-idler pairs labelled $n$ and $-n$.
A graph represents the correlations, where vertices indicate comb frequencies and edges connect signal-idler pairs satisfying the 3-wave mixing criterion. 
We find that all vertices (except vertex 0) have only one incident edge.
These vertices are said to be of degree 1 and correspond to two-mode squeezing.
Note that vertex 0 is unique since it is single-mode squeezed and has an edge to itself, i.e. self-loop.

Going beyond two-mode squeezing require vertices of higher degree and preferably without self-loops. We achieve this with a bichromatic pump, as shown in Fig.~\ref{fig:physics_schematic}c. We partition the entire comb into three infinitely long sub-graphs $K^0$, $K^1$ and $K^{-1}$. 
The superscripts indicate the root vertices, defined as the leftmost vertex in Fig.~\ref{fig:physics_schematic}c. The edges are color-coded to indicate which pump facilitates the interaction.
The graphs for two pumps suggest that a photon detected in any mode could be correlated with an idler at another mode via either the red pump or the blue pump. This ambiguity about exactly which photons are involved in two-mode squeezing can be viewed as an absence of which-color information \cite{Laehteenmaeki2016, Cotler2019}, in analogy to the absence of which-path information required to entangle spatially separated photons \cite{Wang2016}.

We sample the noise output from the flux pumped JPA to construct the measured covariance matrix $\tilde{C}_{nm} = \langle(A_n-\langle A_n\rangle)(A_m - \langle A_m \rangle)\rangle$, where $A_n \in \{I_n, Q_n\}$ is the measured quadrature value for mode $n$.
The measured matrix $\tilde{C}_{nm}$ (obtained in units of \si{\volt\squared}) is scaled according to 
\begin{align}
    \tilde{V}_{nm} = \frac{\tilde{C}_{nm}}{\frac{1}{2} \sqrt{\omega_n \omega_m} Z_c \hbar B},
\end{align}
where $\omega_n$ ($\omega_m$) denotes the frequency of mode $n$ ($m$), while $Z_c$, $B$ and $\hbar$ are the characteristic impedance of the line ($\SI{50}{\Omega}$), the measurement bandwidth ($\SI{1}{kHz}$) and Planck constant respectively.
This scaling permits the covariance matrix to be stated in units of twice the photon number ($2\bar{n}$) and the vacuum state as the identity matrix.
$\tilde{V}_{nm}$ is subsequently compensated for the frequency-dependent gain and added noise of the amplifier chain using a procedure documented in the Supplementary materials.  The result is a calibrated covariance matrix $V'_{nm}$.    

A physical covariance matrix must be semi-positive definite and it must not violate the uncertainty principle.
For a general covariance matrix $V$ in our units, we express these conditions as \cite{Weedbrook2012}
\begin{align}
\begin{matrix}
    V \geq \ 0, && V - i\Omega \geq \ 0,
\end{matrix} \label{eq: constraints}
\end{align}
with the symplectic matrix $\Omega = \bigotimes_{n} \begin{pmatrix} 0 & 1 \\ -1 & 0 \end{pmatrix}$.

When compensating the measured covariance matrix $\tilde{V}_{nm}$ for added measurement noise, it may happen that the calibrated covariance matrix $V'_{nm}$ violates Eqs.~\ref{eq: constraints}.
Therefore, we constrain this compensation to reconstruct a physical covariance matrix, within experimental error $\sigma_{nm}$, that does satisfy Eqs.~\ref{eq: constraints}, by minimizing the objective function \cite{Shchukin_2016}
\begin{align}
    \min_V \left( \max_{nm}\frac{ \left|V'_{nm} - V_{nm} \right|}{\sigma_{nm}} \right). \label{eq:objective function}
\end{align}
Ideally the objective function should be less than unity for a plausible reconstruction (i.e. the reconstructed matrix is within experimental error of the calibrated matrix).
With this reconstructed covariance matrix, we proceed to analyze the entanglement properties.

\begin{figure}[h]
    \centering
    \includegraphics[scale=0.9]{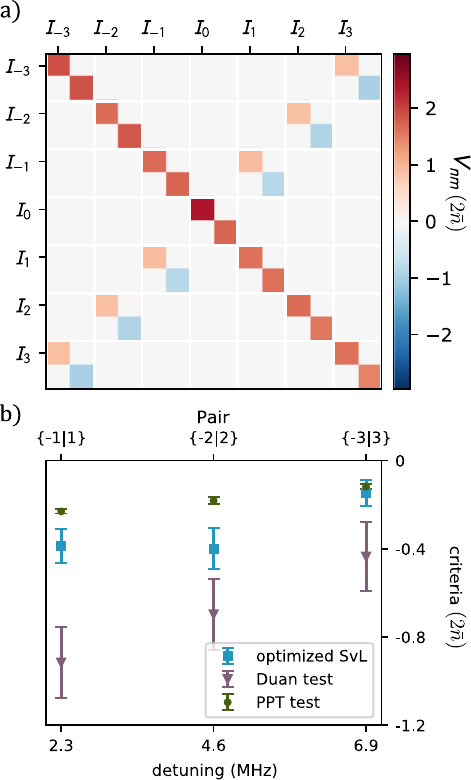}
    \caption{a) Reconstructed covariance matrix generated by a monochromatic pump. The mode at the half-pump frequency is visible at the center where the red diagonal and the red-blue anti-diagonal intersect. The red diagonal contain noise, while the red-blue anti-diagonal indicates pair-wise correlations. For clarity, only the \textit{I}-quadratures are labelled on the x- and y-axes. The  \textit{Q}-quadratures are interleaved between \textit{I} columns and rows. The subscripts label different modes. The covariance matrix is given in units of twice the photon number $2 \bar{n}$.
    \newline
    b) Each mode-pair symmetric around the half-pump frequency \SI{4.3}{GHz} is analyzed for entanglement using the PPT test \cite{Simon2000}, the Duan criterion \cite{Duan2000} and a SvL test~\cite{Shchukin2015, Shchukin_2016} using optimization, see Eq.~\eqref{eq:SvL criteria bipart}.
    With our convention, non-negative values correspond to separable states.
    All three tests return negative results. The bottom x-axis indicates the detuning between mode $n \in [\pm 1, \pm 2, \pm 3]$ and the half-pump frequency mode. Due to our choice of covariance matrix normalization, the criteria are consequently given in units of twice the photon number.
    The errorbars and values are the weighted mean and uncertainties (one standard deviation) from an ensemble of six measurements lasting ten minutes each.
    }
    \label{fig:single_pump_results}
\end{figure}

We first test for entanglement generated by a monochromatic flux pump, (\SI{-3}{dBm} output power) well known to generate pairs of entangled modes, known as two-mode squeezed states \cite{tholen2009, Eichler2011, Beltran2008, Flurin2012, Eichler2014, Wilson2011, Menzel2012, Lahtenmaaki2013, Schneider2020, Perelshtein2021, Esposito2022, Qiu2022}.
We measure the covariance matrix of the closest seven modes around the half-pump frequency with \SI{1}{\kilo\hertz} measurement bandwidth, acquiring $6 \times 10^5$ data points over 10 minutes. To minimize the effect of phase drift in this long-time measurement, we divide the data set into six equal segments.
Reconstructing a physical covariance matrix in each segment,
we find a maximum deviation $\SI{4e-5}{\sigma}$ from the calibrated matrix.
We present one of these covariance matrices in Fig.~\ref{fig:single_pump_results}a.
The 0-mode is located at exactly half the pump frequency.
Consequently, all modes symmetric around the 0-mode create two-mode squeezed states, as indicated by the checkered blue-red anti-diagonal.
The three pairs are analyzed for entanglement using the PPT test \cite{Simon2000} and the Duan criterion \cite{Duan2000}.
The weighted results for the six matrices are presented in Fig.~\ref{fig:single_pump_results}b.
Analysis indicates the three pairs exhibit bipartite entanglement. See the Supplementary material for details.

As depicted in the graphs of  Fig.~\ref{fig:physics_schematic}, a monochromatic pump generates only bipartite states  which are either entangled or separable.  In contrast, a bichromatic pump generates a multipartite state which may belong to one of many possible entanglement and separability classes -- ranging from \textit{fully separable} to \textit{genuinely multipartite entangled} (GME).
As a generalization of the bipartite case, a $n$-partite ($n>2$) mixed state is fully separable if it can be written as a convex combination of product states.
Furthermore, an $n$-partite state can be arranged into $k\le n$ partitions.  If the state is fully separable with respect to $k$-partitioning (or expressed as a mixture of $k$-partitions), we call it \emph{$k$-separable}, if not,  $k$-\emph{inseparable}~\cite{Seevinck2008Sep}.
We call the $n$-partite state \emph{fully inseparable} if it is not separable with respect to any partitioning ~\cite{Dur2000Mar}.  
Additional details are given in the Supplementary material.

To establish whether a $n$-mode state is fully inseparable, it suffices to establish that the state in question is inseparable with respect to any bipartitioning, i.e. it exhibits bi-inseparability.
For pure states, bi-inseparability is a necessary and sufficient condition for GME, as it implies entanglement for any other $k$-partitioning of modes.
However for mixed states, bi-inseparability substantiates only the claim of full inseparability \cite{Teh2019Aug}.
To establish the latter we test for entanglement on all possible bipartitions of the $n$-mode state; a computationally demanding task that grows exponentially with the number of modes $n$ (the number of bipartitions is $2^{n-1}-1$).

\begin{figure*}[t]
    \centering
    \includegraphics[scale=0.75]{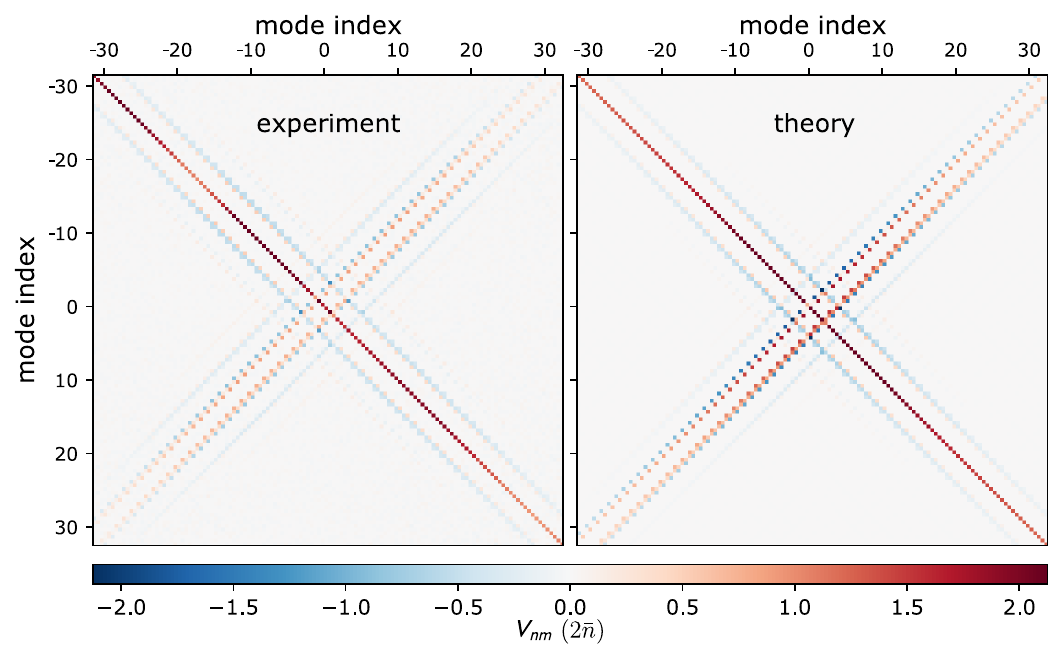}
    \caption{Experimental covariance matrix reconstructed from measurement of 64 modes generated by two pumps (left). The presence of off-diagonals indicate at least the presence of classical correlations. To determine whether these are quantum correlations requires further analysis outlined in the main text. Our theory reproduces a qualitatively similar matrix (right). The finite bandwidth of the JPA weakens correlations among modes at the edges of the frequency comb. For clarity, we omit $I$ and $Q$ labels on the x- and y-axes. Instead they are labelled by the mode index, such that mode 0 lies at exactly half of the average pump frequency, cf. Fig.~\ref{fig:physics_schematic}c. Each mode is spaced by \SI{2.3}{\mega\hertz}. Covariance matrix elements are given in units of twice the photon number ($2 \bar{n}$), making the identity matrix correspond to the vacuum state.}
    \label{fig:complete_cov}
\end{figure*}

We use a bipartition test developed by Shchukin and van Loock \cite{Shchukin2015, Shchukin_2016}, henceforth referred to as the SvL test. Consider the case of a covariance matrix $V$ with zero $IQ$-correlations (or, equivalently, zero $xp$-correlations \cite{Weedbrook2012}). 
This state is equally well characterized by two smaller matrices $V^{II}$ and $V^{QQ}$ containing only $II$- and $QQ$-correlations respectively, a reduction which is always possible with local Gaussian transformations such as single mode phase rotations. Local Gaussian transformations can only transform entangled states into other entangled states, and separable states into other separable states \cite{Duan2000}.
If $n$ modes are divided into bipartitions $\mathcal{I}$ and $\mathcal{J}$, then the SvL condition for separability between $\mathcal{I}$ and $\mathcal{J}$ reads \cite{Shchukin_2016}
\begin{align}
   \mathcal{E} =& \text{Tr}\left[V^{II} (\bm{h} \otimes \bm{h}) \right] + \text{Tr}\left[V^{QQ} (\bm{g} \otimes \bm{g}) \right] \nonumber
   \\ 
   &- 2 |\left\langle h_\mathcal{I}, g_\mathcal{I} \right\rangle| - 2 |\left\langle h_\mathcal{J}, g_\mathcal{J} \right\rangle| \geq 0, \label{eq:SvL criteria bipart}
\end{align}
where $\bm{h}$ and $\bm{g}$ are arbitrary real valued vectors with lengths $n$, and each element is assigned to every mode.
Depending on whether the mode belongs to the $\mathcal{I}$ or $\mathcal{J}$ partition, we extract the corresponding elements from $\bm{h}$ and $\bm{g}$ to create vectors $h_\mathcal{I}$, $g_\mathcal{I}$ and $h_\mathcal{J}$, $g_\mathcal{J}$.
For example, consider three modes $\{1,2,3\}$ and the vectors $\bm{h} = (h_1, h_2, h_3)^\intercal$, $\bm{g} = (g_1, g_2, g_3)^\intercal$.
If the bipartition is $\{\mathcal{I}|\mathcal{J}\}=\{1|2,3\}$, then we have $h_\mathcal{I} = (h_1)$, $g_\mathcal{I}=(g_1)$ and $h_\mathcal{J}=(h_2, h_3)^\intercal$, $g_\mathcal{J}=(g_2, g_3)^\intercal$.

If the bipartition is separable, then the inequality Eq.~\eqref{eq:SvL criteria bipart} holds for all $\bm{h}$ and $\bm{g}$. 
Thus, bi-inseperability is established if we find at least one pair of vectors $\bm{h}$ and $\bm{g}$ which violates the inequality, i.e. for which $\mathcal{E}$ is negative, in each possible bipartition.
To check for consistency with other entanglement tests we apply the SvL test to the monochromatic pump data with the result shown in Fig.~\ref{fig:single_pump_results}b.
The vectors $\bm{h}$ and $\bm{g}$ are optimized to minimize the ratio $\mathcal{E}/\delta\mathcal{E}$, where $\delta\mathcal{E}$ is the uncertainty in $\mathcal{E}$.
From Fig.~\ref{fig:single_pump_results}b, it is clear that all three tests agree on the states' entangled character.

\begin{figure}[h]
    \centering
    \includegraphics[scale=0.8]{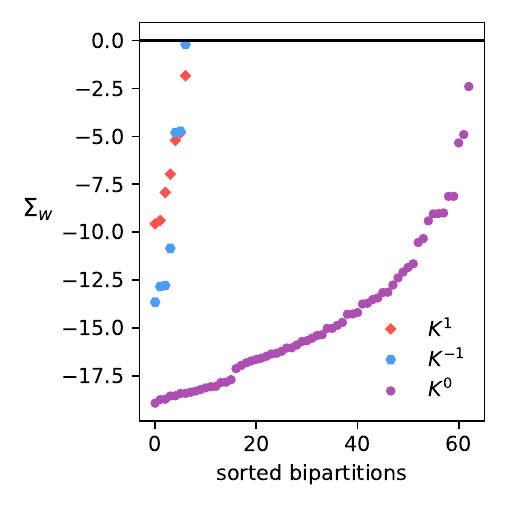}
    \caption{
    Bipartition test for the seven center modes in the $K^0$ set and four center modes from $K^{\pm 1}$. For $K^0$, all tests violate the requirement for separability by at least 2.5 standard deviations, indicating that we have strong evidence for full inseparability. For $K^{1}$, we have full inseparability with a significance of two standard deviations. However evidence for full inseparability for the $K^{-1}$ set is significantly weaker.
    }
    \label{fig:double_pump_results}
\end{figure}

We use the SvL test to demonstrate our main result: the generation of a fully inseparable multipartite state with a bichromatic pump. Figure~\ref{fig:complete_cov} shows the measured covariance matrix of 64 modes  generated by a bichromatic flux-pump each with \SI{-7}{dBm} output power (left).  The measurement is qualitatively well reproduced by our theory (right) provided in the Supplementary material.
With 64 frequencies the number of possible bipartitions exceeds $9 \times {10^{18}}$.
This number is reduced by exploiting the connection topology shown in Fig.~\ref{fig:physics_schematic}.
We divide the covariance matrix into the three smaller subsets of modes: $K^0$, $K^{-1}$ and $K^1$, analyzing each separately and thereby reducing the number of modes to 32 for $K^0$ and 16 modes for $K^{\pm1}$.
But the number of possible bipartitions remains large, roughly $2 \times 10^9$ and $32 767$ respectively.
We therefore analyze an even smaller subset: the seven center modes in the $K^0$-set ($K^0$ modes $\in \{0, \pm2,\pm4, \pm6\}$, and four modes in the $K^{\pm1}$-sets ($K^{-1}$ modes $\in \{-5, -1, 3, 7\}$ and $K^1$ modes $\in \{-7, -3, 1, 5\}$).

As for the monochromatic pump case, we divide a 10 minute measurement with \SI{1}{kHz} bandwidth into six segments, reconstructing the covariance matrix in each, with maximum deviation of $\SI{1.15}{\sigma}$, $\SI{0.95}{\sigma}$ and $\SI{0.92}{\sigma}$ for subsets $K^{-1}$, $K^{0}$ and $K^{1}$ respectively.
We perform the SvL test Eq.~\eqref{eq:SvL criteria bipart} on each matrix of each subset, with vectors $\bm{h}$ and $\bm{g}$ that are individually optimized to minimize $\mathcal{E}/\delta\mathcal{E}$.

The results are summarized in Fig.~\ref{fig:double_pump_results} as a weighted mean entanglement significance $\Sigma_w = \mathcal{E}_w/\delta\mathcal{E}_w$ ($\mathcal{E}_w$ is the mean and $\delta\mathcal{E}_w$ is the corresponding uncertainty).
For example, if $\Sigma_w = -5$, entanglement is established with a statistical significance of five standard deviations.
Since all bipartitions violate the inequality Eq.~\eqref{eq:SvL criteria bipart} by at least 2.5 standard deviations, we have substantial evidence for full inseparability of seven modes.
We probably have full inseparability for four modes in $K^1$, while evidence for $K^{-1}$ is weaker.

In conclusion, direct digital modulation and demodulation methods provide a powerful tool for generation of multipartite entanglement in a microwave frequency comb.  
We demonstrated this with 64 modes using a bichromatic pump and used the SvL test to show full inseparability of a subset of 7 modes in the comb.  
A direct extension of our technique should allow for the creation of multimodal Gaussian cluster states through precise control of the amplitude and phase of the multiple pumps \cite{Arzani2018}. 
For example, pumping at $2\omega_0$ in combination with pumping at the comb spacing $\Delta/2$, should control the scattering between nearest neighbour modes, as in Ref.~\cite{Lecocq2017}. 
Other possible applications include quantum simulation \cite{Hung2021, McDonald2018} and reservoir computing with continuous variable Gaussian states \cite{nokkala2021}.

%TC:ignore
We acknowledge fruitful discussions with Giulia Ferrini and support by the Knut and Alice Wallenberg foundation through the Wallenberg Center for Quantum Technology (WACQT). R.B., M.O.T., and D.B.H. are part owners of the company Intermodulation Products AB, which produces the digital microwave platform used in this experiment.

\bibliography{refs_JPA}
%TC:endignore
\end{document}

\preprint{APS/123-QED}

\title{Supplementary Material: Multipartite entanglement in a microwave frequency comb}

\author{Shan W. Jolin}
\email{shan@meetiqm.com}
\altaffiliation{Present address: IQM Finland Oy, Espoo 02150, Finland}
\affiliation{Department of Applied Physics, KTH Royal Institute of Technology, SE-106 91 Stockholm, Sweden}

\author{Gustav Andersson}
\affiliation{Pritzker School of Molecular Engineering, University of Chicago, Chicago, Illinois 60637, USA}
\affiliation{Department of Microtechnology and Nanoscience MC2, Chalmers University of Technology, SE-412 96 G{\"o}teborg, Sweden}

\author{J. C. Rivera Hern{\'a}ndez}
\affiliation{Department of Applied Physics, KTH Royal Institute of Technology, SE-106 91 Stockholm, Sweden}

\author{Ingrid Strandberg}
\affiliation{Department of Microtechnology and Nanoscience MC2, Chalmers University of Technology, SE-412 96 G{\"o}teborg, Sweden}

\author{Fernando Quijandr{\'i}a}
\altaffiliation{Present address: Quantum Machines Unit, Okinawa Institute of Science and Technology Graduate University, Onna-son, Okinawa 904-0495, Japan}
\affiliation{Department of Microtechnology and Nanoscience MC2, Chalmers University of Technology, SE-412 96 G{\"o}teborg, Sweden}

\author{Jos\'{e} Aumentado}
\affiliation{National Institute of Standards and Technology, 325 Broadway, Boulder, CO 80305, USA}

\author{Riccardo Borgani}
\affiliation{Department of Applied Physics, KTH Royal Institute of Technology, SE-106 91 Stockholm, Sweden}
\affiliation{Intermodulation Products AB, SE-823 93 Segersta, Sweden}

\author{Mats O. Thol{\'e}n}
\affiliation{Department of Applied Physics, KTH Royal Institute of Technology, SE-106 91 Stockholm, Sweden}
\affiliation{Intermodulation Products AB, SE-823 93 Segersta, Sweden}

\author{David B. Haviland}
\affiliation{Department of Applied Physics, KTH Royal Institute of Technology, SE-106 91 Stockholm, Sweden}
\email{haviland@kth.se}

\date{\today}

\maketitle

\section{Measurement setup}

\begin{figure*}
    \centering
    \includegraphics[scale=0.75]{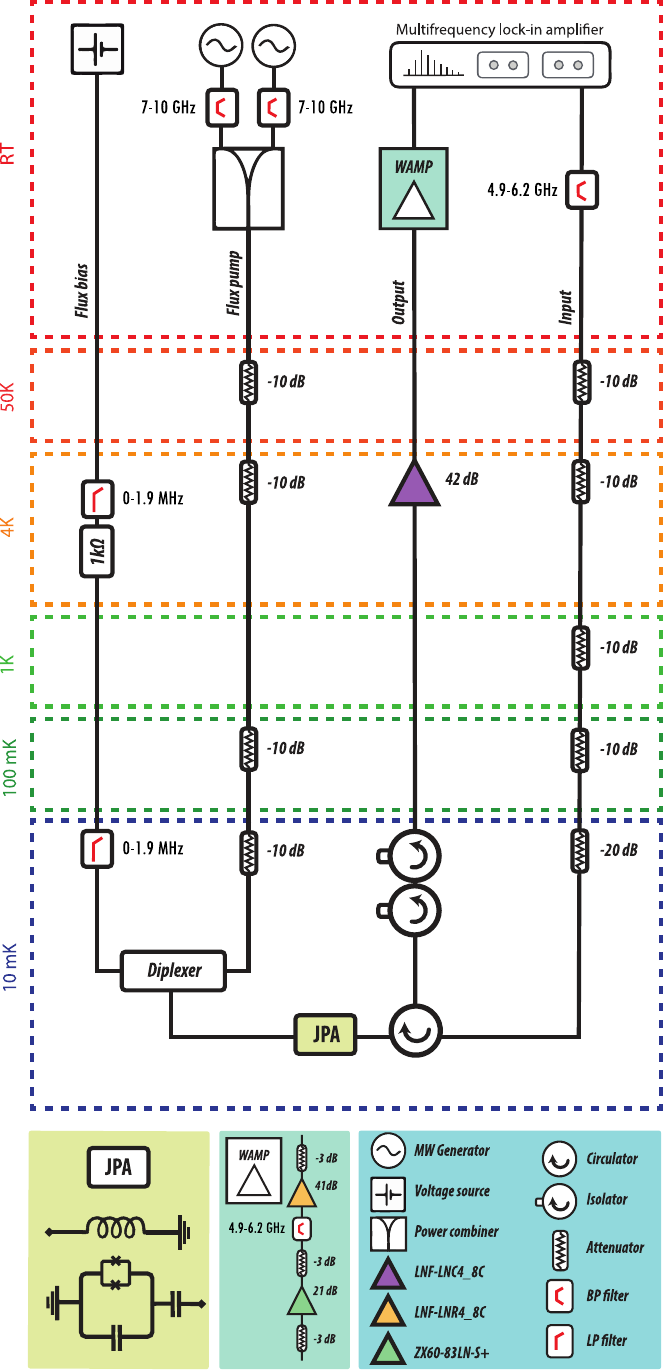}
    \caption{Schematic of complete measurement setup including cryogenic electronics. Note that despite the JPA signal input is connected to our multi-frequency lockin, it was not used in the measurements presented in the main text. A signal was provided only for characterization purposes.
    Here we use a bandpass filter in the warm amplifier (WAMP) stage to \SI{4.9}{}--\SI{6.2}{GHz} range with the purpose to reduce aliasing.
    }
   \label{fig:fridge_setup}
\end{figure*}

The Josephson parametric amplifier (JPA) is cooled to $\SI{10}{mK}$ with a dilution refrigerator and operated in reflection, which means that the input and output share the same port. The two opposite travelling waves are separated by a circulator (see Fig.~1a) permitting us to further amplify only the outgoing signal. A pair of isolators prevents noise from the following amplifier from reaching the JPA signal port.

An RF pump applied at a second port modulates the flux in the JPA SQUID. We also apply a DC current through the pump port to tune the JPA resonance down to $\SI{4.2}{GHz}$.
A diplexer at the \SI{10}{\milli\kelvin}-stage of the dilution refrigerator combines the DC current and RF pumps.
See Fig.~\ref{fig:fridge_setup} for schematic description of the entire measurement setup.  We pump at twice the JPA resonance frequency using external signal generators locked to the measurement sampling clock.

Measurement of the noise covariance is performed with Vivace from Intermodulation Products \cite{intermodulation_products} running a continuous wave firmware which implements a multifrequency lockin amplifier.  The lockin is designed to work with orthogonal frequencies on the time-window used for demodulation.  A process refered to as tuning  fixes the drive and measurement frequencies such that they are integer multiples of the measurement bandwidth (inverse of the time window) while being commensurate with the sampling frequency \cite{tholenRSI_2011, imp_manual}. 

Based on the latest Radio Frequency System-on-a-Chip (RFSoC) technology, Vivace uses fast Digital-Analog Converters (DAC) at \SI{5}{GSamples/s}, resulting in a Nyquist frequency of \SI{2.5}{\giga\hertz}. We operate the DAC in the second Nyquist zone to directly access the microwave noise.
Bandpass filters select the noise in the second Nyquist zone, measured at its alias frequency in the 1st zone, while rejecting noise at that frequency in the 1st zone.  This digital method of down conversion eliminates analog mixers, thus removing another source of error from improperly calibrated mixers \cite{Jolin2020}.
We use a bandpass filter at the warm amplifier stage to prevent aliasing when sampling the noise.
However the filter (VBFZ-5500-S+) has a specified range of \SI{4.9}{}--\SI{6.2}{\giga\hertz}, meaning that measurements at \SI{4.3}{\giga\hertz} suffer from additional loss.

\section{Calibration}

To reconstruct the quantum state before amplification, we need an accurate estimate of the gain and added noise of the entire amplification chain.
If we consider the amplification chain as a noisy bosonic Gaussian channel \cite{Weedbrook2012} where each mode, labelled by index $n$, is subject to a frequency-dependent gain $G_n$ and an average number of added noise photons $\bar{n}_n$.  Amplification transforms the multimodal quantum covariance matrix $V'$ into the measured covariance matrix $\tilde{V}$ according to
\begin{align}
    \tilde{V} = T V' T^\top + N,
    \label{eq: covtransformation}
\end{align}
where $T = \bigoplus_{n} \sqrt{G_n} I_n$ and $N = \bigoplus_{n} (G_n-1)(2 \bar{n}_n+1)I$.
Here $\bigoplus$ is the direct sum and $I$ the identity matrix.
By inverting $T$, the quantum covariance matrix $V'$ is reconstructed according to
\begin{align}
    V' = T^{-1}\left(\tilde{V} - N \right)(T^\top)^{-1},
    \label{eq: inverse covtransformation}
\end{align}
using calibrated values for $G_n$ and $\bar{n}_n$.
All our expressions assume covariance matrices are normalized such that the vacuum state corresponds to the identity matrix.

We estimate parameters $G_n$ and $\bar{n}_n$ by measuring the noise emitted from a matched resistor $R = \SI{50}{\Omega}$ as a function of the temperature $T$, a method known as Planck spectroscopy \cite{Mariantoni2010, Tholen_thesis}. This black-body noise is reflected off of the quiescent JPA, operated with no flux pump or flux bias.  There are two noise sources: the \SI{-20}{dB} attenuator on the input signal line to the JPA and the matched load of the isolators, both at the \SI{10}{\milli\kelvin}-stage.
Their temperature is controlled by slowly heating up the entire \SI{10}{\milli\kelvin}-stage, allowing the temperature to stabilize for \SI{20}{min} before measuring.

\begin{figure*}
    \centering
    \includegraphics[scale=0.7]{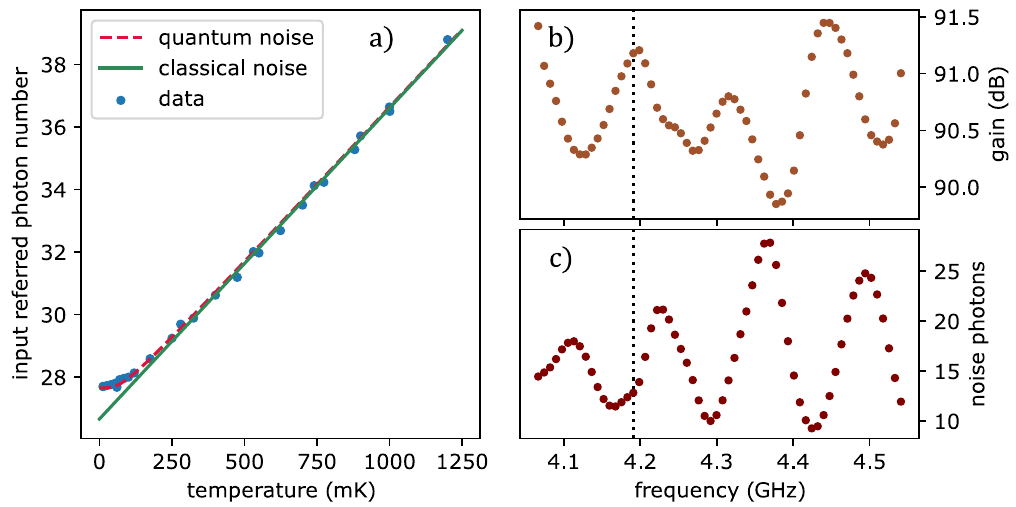}
    \caption{
    \textbf{a}: Number of photons referred to the amplifier input. The fit provides with the values $G$ (gain) and $\Bar{n}$ (added noise) at the frequency specified by the dashed vertical line in figures b and c (at \SI{4.19}{GHz}). The deviation from the green line is due to vacuum fluctuations.
    \textbf{b} and \textbf{c}: The gain and added noise is plotted as a function of frequency. 
    In all figures, the errorbars indicating one standard deviation are smaller than the marker size.
    }
    \label{fig:calibration_curve}
\end{figure*}

The power spectral density of the noise is given by \cite{Clerk2010}
\begin{align}
    P &= \frac{\hbar \omega}{2}\coth{\frac{\hbar \omega}{2 k T}}. \label{eq:power_spectral_density}
\end{align}
An amplifier scales this noise power by the gain $G$ and adds an average number of noise photons $\Bar{n}$,
\begin{align}
    P &= G \hbar \omega  \left[ \frac{1}{2}\coth{\frac{\hbar \omega}{2 k T}} + \frac{1}{2}(1 + 2\Bar{n} ) \right]. \label{eq: power_spectral_density_fit}
\end{align}
In terms of the voltage variance over a measurement bandwidth $\Delta_f$ this expression reads
\begin{align}
    \langle V^2 \rangle &= 4 \Delta_f G \hbar \omega R \left[ \frac{1}{2}\coth{\frac{\hbar \omega}{2 k T}} + \frac{1}{2}(1 + 2\Bar{n} ) \right]. \label{eq: voltage_variance}
\end{align}
Note that even if the added noise $\Bar{n} = 0$, there is always a half-photon of added noise power, which reflects the nature of phase insensitive amplification \cite{Caves1982}.
Figure~\ref{fig:calibration_curve}a depicts the fit of Eq.~\eqref{eq: power_spectral_density_fit} to the noise power, there expressed instead in units of photon number at the amplifier input (hence this quantity is divided by the amplifier gain) called the input referred photon number.

The noise power for the calibration is obtained at all frequencies simultaneously (in the same time window) with the multifrequency lockin. 
The data (Fig.~\ref{fig:calibration_curve}a) clearly shows the flattening at lower temperatures, revealing the measurement of quantum fluctuations.
The extracted gain $G$ and $\Bar{n}$ as a function of frequency is shown in Fig.~\ref{fig:calibration_curve}b and Fig.~\ref{fig:calibration_curve}c respectively. 
Heating the \SI{10}{\milli\kelvin}-stage in this manner places the reference plane for the calibration (as seen by the amplifier) close to the first isolator and after the JPA. 
A better calibration procedure that takes in to account the small insertion losses of the isolator and circulator, is expected to improve the purity of our reconstructed data.

\section{Josephson parametric amplifier}
\begin{figure}
    \centering
    \includegraphics[scale=0.4]{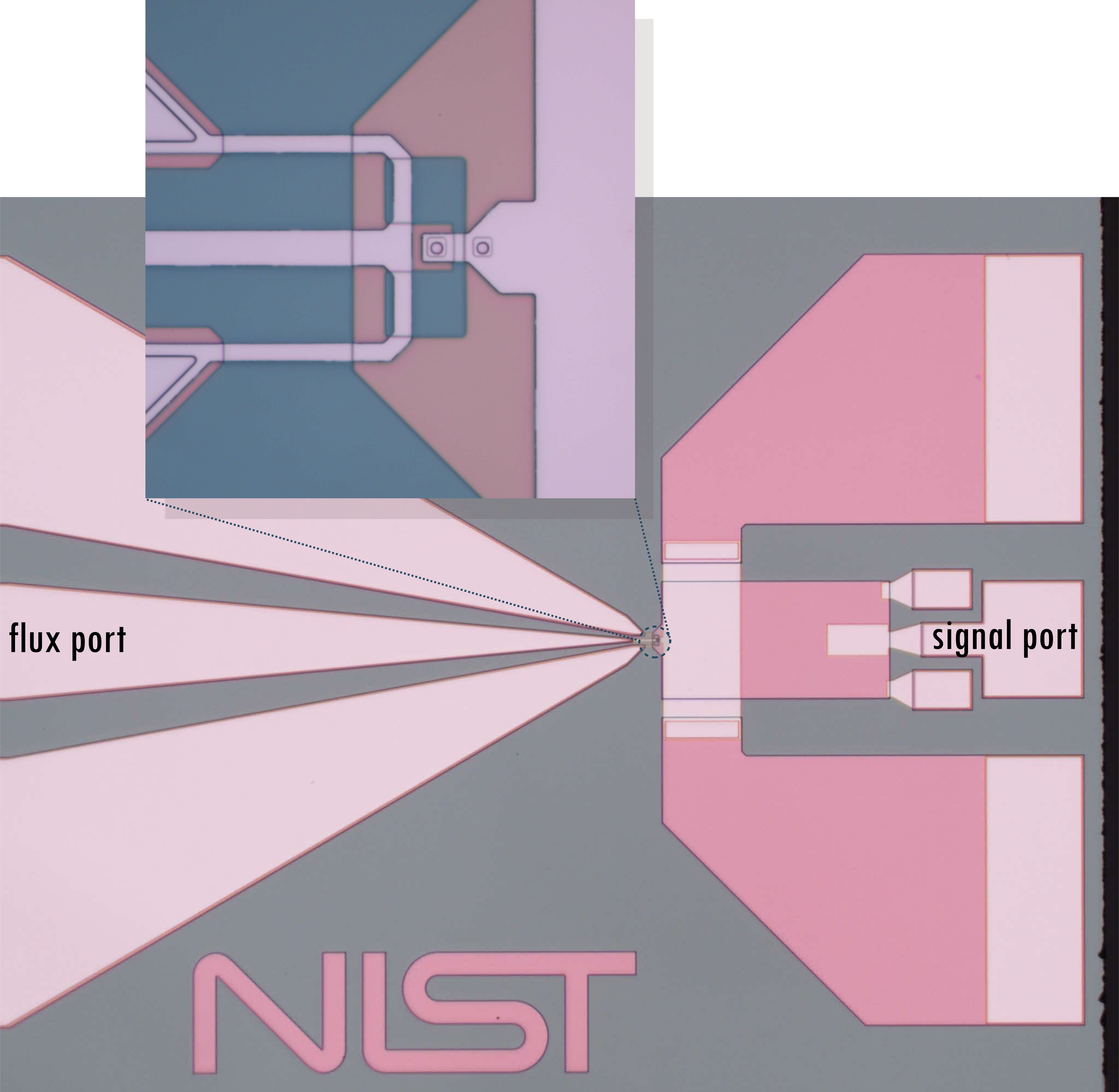}
    \caption{
    An optical image of the JPA device with an inset zoomed in on the gradiometric dc-SQUID. Flux and signal parts are labelled. The width of the chip is about \SI{5}{mm}.
    }
    \label{fig:jpa_image}
\end{figure}

The JPA is fabricated in the NIST Boulder cleanroom using an optical lithography process for multi-layer superconducting circuits with niobium trilayer junctions. The layer stack details are given in the appendix of Ref.~\cite{Lecocq2017}. 
The circuit, shown in Fig.~\ref{fig:jpa_image}, has a parallel $LC$ resonance, where $C \approx 12\,pF $ is realized by an overlap capacitor with low-loss amorphous silicon as the dielectric ($\epsilon\approx9$, $\tan \delta \sim$1.5--5$\times 10^{-5}$).  $L$ is the Josephson inductance of a gradiometric dc-SQUID, with junction critical currents, $I_0 \simeq 5\,\mu$A, giving a total SQUID critical current, $I_\mathrm{SQ} \simeq 10\,\mu$A. The SQUID's Josephson inductance is modulated with external magnetic flux from a flux line.  The flux port is driven by a diplexer (see fig. \ref{fig:fridge_setup}) for both DC bias and AC pumping. The circuit was designed for a maximum resonant frequency of approximately 8\,GHz. At zero flux bias the resonance is found at \SI{7.8}{GHz}.
The resonator is coupled to the signal port through a capacitance $C_c\approx 0.5$\,pF which fixes the loaded bandwidth of the JPA.

For entanglement measurements we bias the JPA resonance with DC flux to achieve $\omega_0/2\pi\approx 4.3$\,GHz ($\Phi_\mathrm{dc}\simeq \pm0.4 \Phi_0$), where we measure a loaded bandwidth of $\kappa\approx124$\,MHz.  At this flux bias, where the slope of the resonance frequency versus flux is roughly linear, we modulate (pump) the flux at $\Omega_{1,2}\sim 2\omega_0$ to generate 3-wave mixing (in contrast to 4-wave mixing or `doubly-degenerate' pumping where $\Omega_{1,2} \sim \omega_0$). Three-wave mixing has the advantage that pump-to-signal port leakage, which may interfere with measurements of multi-mode entanglement, is far from the mode frequencies of interest.

\begin{figure}
    \centering
    \includegraphics[scale=0.7]{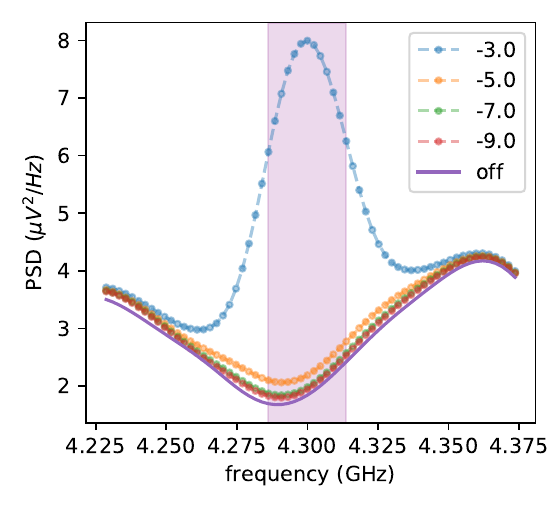}
    \caption{
    The noise PSD from a bichromatically pumped JPA. The different colored lines indicate the pump power of each pump in dBm (both pumps are equal in power). For the entanglement tests, the JPA was operated at \SI{-7}{dBm}. The shaded region indicates the frequency range where the modes for our entanglement test reside.}
    \label{fig:jpa_psd}
\end{figure}

For all entanglement tests we operate the JPA at relatively low pump power.
For the monochromatic case the applied output pump power at the signal generator output is \SI{-3}{dBm}.
With a bichromatic pump however, we operate the JPA with each pump at \SI{-7}{dBm}.
Fig.~\ref{fig:jpa_psd} shows the power dependence of the noise power spectral density (PSD) emitted by the JPA subjected to a bichromatic pump.
At \SI{-3}{dBm} a Lorentzian-like shape appears, indicating gain in the JPA.  At \SI{-7}{dBm} we see that the PSD lifts slightly above the pump-off reference (solid line). We found that \SI{-3}{dBm} added too much noise to the data for investigation of multi-modal entanglement, but \SI{-7}{dBm} was a good compromise between strong correlations and low added noise.
The shaded region indicates the frequency range where the entangled modes from the main text can be found.

\section{Covariance Matrix Reconstruction}

\subsection{Error propagation}
The experimental errors in the covariance matrix $V'$ should take into account uncertainties in both the measurement and the calibration.
We estimate measurement errors by applying a bootstrap method (see \cite{Pastore2019} for a pedagogical discussion) to each element in the covariance matrix.
The two sources of error are appropriately combined through error propagation based on Eq.~\eqref{eq: inverse covtransformation} in the main text (assuming $G \gg 1$), to obtain an estimate of the uncertainty of each element in $V'$, expressed as an error matrix $\sigma_{nm}$.
\begin{align}
    \sigma^2_{nm} = 
    \begin{cases} \left( \frac{\sigma^G_{n} \tilde{V}_{nn}}{G_n^2} \right)^2 + \left(2\sigma^{\Bar{n}}_n\right)^2 + 4\frac{\tilde{V}_{nn} \sigma^{G{\Bar{n}}}_i}{G_i^2} + \left( \frac{ \sigma_{nm}^{\tilde{V}} }{ G_n } \right)^2 \\ \quad \quad \mbox{ if } n=m,
    \\ 
    \left( \frac{\sigma_n^G \tilde{V}_{nm}}{2 \sqrt{G_n^3 G_m}}\right)^2 + \left( \frac{\sigma_m^G \tilde{V}_{mn}}{ 2 \sqrt{G_m^3 G_n} } \right)^2 +
    \left( \frac{ \sigma_{nm}^{\tilde{V}} }{ \sqrt{G_n G_m}}\right)^2 \\ \quad \quad \mbox{ if } n \neq m, \end{cases}
\end{align}
where $\sigma^G_{n}$ and $\sigma^{\Bar{n}}_{n}$ are uncertainties in the parameters $G$ and $\Bar{n}$. 
As we cannot assume these are completely independent, their covariance $\sigma^{G{\Bar{n}}}_n$ has to be accounted for. The measurement errors in ${\tilde{V}_{nm}}$ are denoted $\sigma^{\tilde{V}}_{nm}$.

For the SvL criteria Eq.~(4), the uncertainty in the quantity $\mathcal{E}$ is estimated using error propagation, for which the result is
\begin{align}
    \delta\mathcal{E} = \sqrt{ \sum_{\alpha\beta} \left( {(\sigma^{II})}^2_{\alpha\beta} h^2_\alpha h^2_\beta + { (\sigma^{QQ})}^2_{\alpha\beta} g^2_\alpha g^2_\beta \right) }. \label{eq:error entanglement}
\end{align}
The uncertainties in the covariance matrix are denoted by $\sigma^{II}$ and $\sigma^{QQ}$ (corresponds to the uncertainties in elements of $V^{II}$ and $V^{QQ}$ respectively, see the main text).

\subsection{Additional details on data analysis}

\begin{figure*}
    \centering
    \includegraphics[scale=0.7]{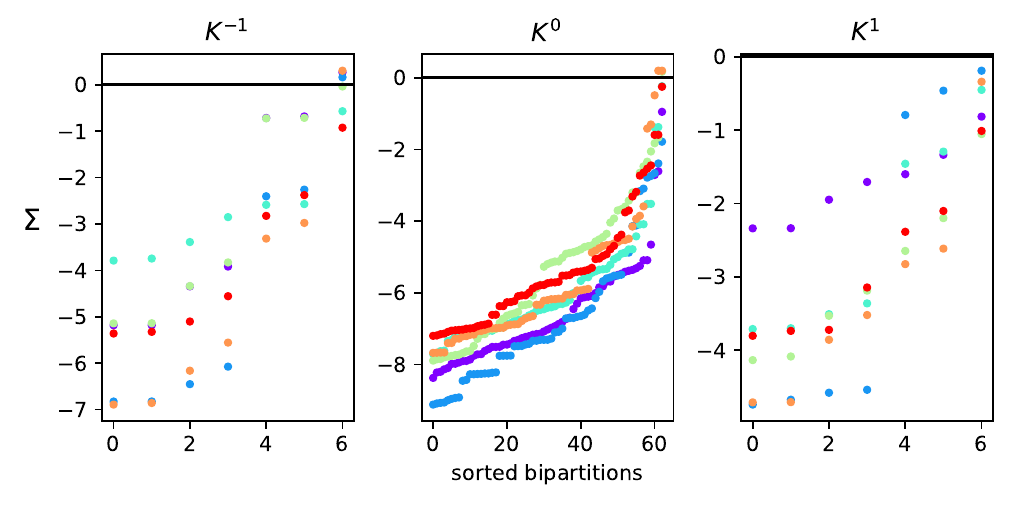}
    \caption{
    We plot the SvL bipartition test result for the six covariance matrices in three mode subsets. Points with the same color stem from the same matrix. The weighted mean of points belonging to the same bipartition are the $\Sigma_w$ points in the main text Fig.~4.
    }
    \label{fig:bipart_all_results}
\end{figure*}

The optimization of Eq.~(3) is performed by convex optimization \cite{agrawal2018, diamond2016}.
As mentioned in the main text, for each of the three subsets of independent modes, $K^{-1}$, $K^{0}$, $K^{1}$, we analyze six covariance matrices from which a maximum value of 1.15$\sigma$, 0.95$\sigma$ and 0.92$\sigma$ standard deviations is obtained for each subset respectively.
Each of the six matrices in each subset is tested for entanglement using Eq.~(4) from the main text.
Optimization is in this case carried out using differential evolution \cite{diff_evolution}.
The results of these individual tests are shown in Fig.~\ref{fig:bipart_all_results}.
Figure~4 in the main text presents the weighted mean for each subset, which is calculated according to 
\begin{align}
\renewcommand{\arraystretch}{2}
\begin{matrix}
    \mathcal{E}_w = \frac{ \sum\limits^6_{k=1}w_k\mathcal{E}_k} {\sum\limits^6_{k=1}w_k}, &&
    \delta\mathcal{E}_w =\frac{1}{\sqrt{\sum\limits^6_{k=1}w_k}}, &&
    w_k = \frac{1}{\delta\mathcal{E}_k^2}.
\end{matrix}
\end{align}

\section{Covariance Matrix of Output Noise}

\subsection{JPA Hamiltonian}\label{sect:main}

The Hamiltonian for a resonator terminated in a Josephson junction is
\begin{align}
H = \hbar\omega_r \,a^\dagger a - E_J \cos \left( \frac{\Phi}{\varphi_0}  \right),
\end{align}
where $\Phi$ is the total flux across the junction and $\varphi_0 = \hbar / 2e$ is the reduced magnetic flux quantum.
We expand the cosine up to second-order in $\Phi/ \varphi_0$
\begin{align}
\cos \left( \frac{\Phi}{\varphi_0}  \right)  = 1 - \frac{1}{2!}  \left( \frac{\Phi}{\varphi_0}  \right)^2 + \mathcal O\left[ \left( \frac{\Phi}{\varphi_0}  \right)^3  \right] .
\end{align}
Discarding the constant term, we have
\begin{align}
H = \hbar\omega_r \,a^\dagger a + \frac{E_J}{2!} \left( \frac{\Phi}{\varphi_0}  \right)^2 .
\end{align}
Now,  the flux field $\Phi$ is given by
\begin{align}\label{eq:flux-field}
\Phi = \phi_{\rm ZPF} (a^\dagger + a),
\end{align}
where $\phi_{\rm ZPF}$ 
correspond to the zero-point fluctuations (ZPF) in the flux.
Then,  we finally get for the Hamiltonian
\begin{align}
H = \hbar\omega_r\, a^\dagger a + \frac{E_J}{2!} \phi_0^2 (a^\dagger + a)^2 ,
\end{align}
where we have introduced the reduced ZPF $\phi_0 \equiv \phi_{\rm ZPF} / \varphi_0$.
This can be further simplified to
\begin{align}
H = \hbar\widetilde \omega_r\, a^\dagger a + \frac{E_J}{2!} \phi_0^2 (a^{\dagger 2} + a^2 )
\end{align}
by introducing the renormalized resonator frequency $\widetilde \omega_r = \omega_r + E_J \phi_0^2/\hbar$.

Replacing the Josephson junction by a DC SQUID,  we gain tunability of the Josephson energy: $E_J \to E_J(\phi_{\rm ext}) = E_{J, 0} \cos (\phi_{\rm ext})$.
Consider the following time-dependent external magnetic flux
\begin{align}
\phi_{\rm ext}(t) = \phi_{\rm dc} + \phi_{\rm ac}(t) ,
\end{align}
where $ \phi_{\rm dc}$ is a time-independent bias and $\phi_{\rm ac}(t)$ is a small amplitude modulation supplied by a flux pump. The amplitude is small in the sense that $\vert \phi_{\rm ac}(t) /\phi_{\rm dc}  \vert \ll 1$.
In this regime we have different order approximations of the Josephson energy in the modulation amplitude.
To first-order 
\begin{align}
\cos \phi_{\rm ext}(t) &= \cos \left( \phi_{\rm dc} + \phi_{\rm ac}(t) \right) \nonumber \\
&= \cos \phi_{\rm dc} - \phi_{\rm ac}(t) \sin \phi_{\rm dc},  \label{eq:first-order}
\end{align}
and second-order 
\begin{align}
\cos \phi_{\rm ext}(t) &= \cos \left( \phi_{\rm dc} + \phi_{\rm ac}(t) \right) \nonumber \\
&= \cos \phi_{\rm dc} \cos \phi_{\rm ac}(t)  -  \sin \phi_{\rm dc} \sin \phi_{\rm ac}(t) \nonumber \\
&=  \cos \phi_{\rm dc} \left( 1 - \frac{1}{2!} \phi_{\rm ac}(t)^2  \right) -  \phi_{\rm ac}(t)  \sin \phi_{\rm dc} \nonumber  \\
&=  \cos \phi_{\rm dc} - \left(  \phi_{\rm ac}(t)  \sin \phi_{\rm dc} + \frac{1}{2!} \phi_{\rm ac}(t)^2   \cos \phi_{\rm dc}    \right).\label{eq:second-order}
\end{align}
We find it sufficient to consider only up to first-order to qualitatively reproduce the features in our data. Notice that if we had not applied any static flux bias, i.e. $\phi_{\rm dc}=0$, then the first-order expression Eq.~\eqref{eq:first-order} would reduce to zero and the second-order expression Eq.~\eqref{eq:second-order} would be necessary.

In this work we consider a modulation of the form
\begin{align}\label{eq:freq-comb}
\phi_{\rm ac}(t) = \sum_{n=1}^{n_s} i \lambda_n \exp \left[- i(2\widetilde{\omega}_r + \delta_n)t \right] + {\rm h.c.} ,
\end{align}
that is, a sum of $n_s$ sinusoidal waves at near twice the resonator frequency. 
For a bichromatic pump, $n_s=2$.
Here $\lambda_n$ denote the complex amplitudes and $\vert \delta_n  \vert \ll \widetilde{\omega}_r$ the small detunings from $2\tilde{\omega}_r$.

In the first-order approximation Eq.~\eqref{eq:first-order}, the Hamiltonian for the resonator and SQUID becomes
\begin{align}
H \approx \, &\hbar\widetilde{\omega}_r\,  a^\dagger a + \frac{1}{2!} (E_{J,0} \cos \phi_{\rm dc}) \phi_0^2 \left( a^{\dagger 2} + a^2 \right)\nonumber \\ &- \frac{1}{2!} (E_{J,0}  \sin \phi_{\rm dc}) \phi_0^2 \phi_{\rm ac}(t) \left( a^{\dagger} + a \right)^2 .
\end{align}
Note that for the time dependent part we are considering the full $\Phi^2$ 
potential~[cf. Eq.~\eqref{eq:flux-field}].
As customary,  in order to study the slow dynamics resulting from the modulation, we transform the Hamiltonian 
into the rotating frame defined by $R(t) = \exp(i \widetilde{\omega}_r a^\dagger a \,t)$, i.e.,
$H \to H_{R} = R H R^\dagger + i \dot{R} R^\dagger$:
\begin{align}
H_{R}(t) \approx &\frac{1}{2!} (E_{J,0} \cos \phi_{\rm dc}) \phi_0^2 \left( a^{\dagger 2} {\rm e}^{+2 i \widetilde{\omega}_r t} + {\rm h.c.} \right)  \nonumber\\
&-  \frac{1}{2!} (E_{J,0}  \sin \phi_{\rm dc}) \phi_0^2 \sum_n \left[ i \lambda_n {\rm e}^{- i(2\widetilde{\omega}_r + \delta_n)t } + {\rm h.c.}  \right] \nonumber \\ &\times \left( a^{\dagger 2} {\rm e}^{+2 i \widetilde{\omega}_r t} + a^{2} {\rm e}^{-2 i \widetilde{\omega}_r t} + 2 a^\dagger a \right) .
\end{align}
In the rotating wave approximation (RWA) we keep only the slow rotating terms, i.e., those rotating at frequencies much smaller than $\widetilde{\omega}_r$
\begin{align}\label{eq:Hslow-first}
H_{\rm RWA} \approx -\frac{i}{2!} (E_{J,0}  \sin \phi_{\rm dc}) \phi_0^2 \sum_{n=1}^2  \lambda_n a^{\dagger 2}  {\rm e}^{- i \delta_n t}  + {\rm h.c. }
\end{align}

\subsection{Langevin equation of motion}

Introducing the pump frequencies $\omega_{p,i} = 2 \widetilde{\omega}_r + \delta_i$, $i = 1,2$, we can re-write the first-order RWA Hamiltonian \eqref{eq:Hslow-first} as
\begin{align}
H_{\rm RWA} = -\frac{i\hbar}{2} \sum_{n=1}^2 \left[ \mu_n^* a^2 {\rm e}^{+ i( \omega_{p,n} - 2\widetilde{\omega}_r)t} - \mu_n a^{\dagger 2} {\rm e}^{-i (\omega_{p,n} - 2 \widetilde{\omega}_r) t}  \right],
\end{align}
where we introduce $\mu_n = \left(E_{J,0}\sin\phi_{\rm dc}\right)\phi_0^2\lambda_n/\hbar$ for notational brevity.
Now we un-do the rotation transformation $R(t)$ which leads us to
\begin{align}
H \approx \hbar\widetilde{\omega}_r a^\dagger a -\frac{i\hbar}{2} \sum_{n=1}^2 \left( \mu_n^* a^2 {\rm e}^{+ i \omega_{p,n} t} - \mu_n a^{\dagger 2} {\rm e}^{-i \omega_{p,n}  t}  \right) .
\end{align}
From here  we derive the Heisenberg equation of motion $\partial_t \, a = i[H,  a] / \hbar$. For convenience, we work in the frequency domain defined by
\begin{align}
a[\omega] = \int_{-\infty}^{+\infty} {\rm d}t \, {\rm e}^{i \omega t}\,  a(t) \label{eq:Fourier1}.
\end{align}
Therefore, the Langevin equation of motion is
\begin{align}\label{eq:Langevin-full}
\left[ \frac{\kappa}{2} - i (\omega - \widetilde{\omega}_r )  \right] a[\omega] -\sum_{n=1}^2 \mu_n a^\dagger [ \omega_{p,n} - \omega] = - \sqrt{\kappa} a_{\rm in} [\omega] ,
\end{align}
where we included the presence of the environment with the external decay rate $\kappa$ of the resonator (ignoring internal losses) and $a_{\rm in}$ the input operator \cite{Gardiner1985}.
This equation tells us that for every "signal" frequency $\omega$ there are two "idlers"
 with frequencies $\omega_{p,1} - \omega$ and $\omega_{p,2} - \omega$.

\subsection{Frequency comb output correlations}\label{sec: cov matrix theory}

The idea now is to characterize the output correlations resulting from the mixing of equally spaced input signal frequencies, i.e., a \textit{frequency comb}.
This follows from studying the output frequency components given by the input-output equation
$a_{\rm out}[\omega] = a_{\rm in}[\omega] + \sqrt{\kappa} a[\omega] $~\cite{Gardiner1985}. In order to do this we need to first define the frequency comb.

We start by introducing the following notation
$\delta_1 = -\delta_2  \equiv -\Delta/2$.  In terms of $\Delta$ the pump frequencies can be re-written as
\begin{align}
\omega_{p,1} &=  2\widetilde{\omega}_r - \Delta/2 \nonumber \\
\omega_{p,2} &=  2\widetilde{\omega}_r + \Delta/2. \nonumber
\end{align}
Without loss of generality we assume  $\delta >0$ and therefore $\omega_{p,2} > \omega_{p,1}$.

The frequency comb is defined as an equally spaced set of frequencies labelled by an integer index $n$
\begin{align}
\omega_{n} = \omega_0 + n \,\delta . \nonumber
\end{align}
We choose the center of the comb as
 $\omega_0 \equiv (\omega_{p,1} +\omega_{p,2})/4  =\widetilde{\omega}_r $ and its spacing $\delta$ as
\begin{align}
\delta &\equiv \frac{\omega_{p,2}}{2} - \omega_0 = \omega_0 -\frac{\omega_{p,1}}{2} \nonumber\\
&= \frac{\omega_{p,2} -\omega_{p,1}}{4} = \frac{\Delta}{4}.
\end{align}

Following Eq.~\eqref{eq:Langevin-full},
for a signal $\omega$ in the frequency comb, i.e., $\omega = \omega_n$,
the two idlers are located at $\omega_{p,1} - \omega_n$ and 
$\omega_{p,2} - \omega_n$. We can rewrite these as
\begin{align}
\omega_{p,1} - \omega_n &= 2 \omega_0 - 2\delta - (\omega_0 + n \delta) \nonumber\\
&= \omega_0 - (n+2) \delta \nonumber\\
&= \omega_{-n-2} \nonumber
\end{align}
and
\begin{align}
\omega_{p,2} - \omega_n &= 2 \omega_0 + 2\delta - (\omega_0 + n \delta) \nonumber\\
&= \omega_0 - (n-2) \delta \nonumber\\
&= \omega_{-n+2} ,\nonumber
\end{align}
in other words the idlers are also within the frequency comb with an index spacing of $\pm 2$.
Therefore, if $n$ is an even (odd) integer,  it will only be coupled to other even (odd) frequencies, or equivalently, there will be \textit{no correlations} between even and odd indices in the frequency comb.
Furthermore,  while the two pumps connect all even modes with each other, the odd ones are split into two subsets:
\begin{align}
\{ \dots -11, -7, -3, +1, +5, +9, \dots  \} \nonumber\\
\{ \dots -9, -5, -1, +3, +7, +11, \dots  \} \nonumber
\end{align}
See Fig.~1 in the main text for a visual rendition of the correlation across modes in the comb.

Finally,  by introducing the  notation $a[\omega_n] \equiv a_n$ for the frequency-domain operators, we rewrite the Langevin equation \eqref{eq:Langevin-full} for the frequency comb as:
\begin{align}\label{eq:Langevin-comb}
\chi^{-1}_n a_n - \mu_1 a^\dagger_{-2-n} - \mu_2 a^\dagger_{-n+2}= - \sqrt{\kappa} a^{\rm in}_n ,
\end{align}
with $\chi_n = [\kappa/2 - i(\omega_n - \widetilde{\omega}_r)]^{-1}$.

In order to solve for $a_{\rm out}[\omega_n] \equiv a_{{\rm out}, n}$, we can write the system \eqref{eq:Langevin-comb} in matrix form
\begin{align}\label{eq:Langevin-matrix}
M \bm{a} = -\sqrt{\kappa} \bm{a}_{\rm in} ,
\end{align}
with $\bm{a}^\top = (\dots, a_{-1}, a_{0}, a_{+1},  \dots ,  a^\dagger_{-1}, a^\dagger_{0}, a^\dagger_{+1}, \dots )$ and similarly for $\bm{a}_{\rm in}$ and $\bm{a}_{\rm out}$. From the above equation
we have $\bm{a} = -\sqrt{\kappa}M^{-1} \bm{a}_{\rm in}$.
The solution for the output frequency components follows from the input-output relation
\begin{align}
\bm{a}_{\rm out} &= \bm{a}_{\rm in} + \sqrt{\kappa} \bm{a} \nonumber \\
&= (\mathbb{I} + \kappa M^{-1}) \bm{a}_{\rm in} =  S  \bm{a}_{\rm in} . \label{eq:S-matrix}
\end{align}
The quadrature operators $x_{ n} = b^\dagger_n + b_n$ and  
$p_n = i(b^\dagger_n - b_n)$ can be obtained via the linear transformation $\bm{r}  = K \bm{b}$ with $\bm{b} = \bm{a},\, \bm{a}_{\rm in}, \, \bm{a}_{\rm out}$ and the matrix $K$ given by
\begin{align}\label{eq:K-matrix}
K =\begin{pmatrix}
\mathbb{I}_n & \mathbb{I}_n \\
-i\mathbb{I}_n & i\mathbb{I}_n
\end{pmatrix} ,
\end{align}
when the quadratures are arranged in the $(x,p)$ ordering 
$\bm{r} = (x_1, \dots, x_n,\, p_1, \dots, p_n)$.

The linear relation between input and output modes \eqref{eq:S-matrix} is a consequence of having neglected the non-linear contributions from the SQUID potential.  
This linear transformation will convert an input thermal (Gaussian) state into an output Gaussian state as well.
A Gaussian state is completely characterized by its first and second order moments, the latter being known as the covariance matrix $V$. The covariance matrix is defined as
\begin{align}
V = \frac{1}{2} \langle \bm r \cdot \bm r^\top + ( \bm r \cdot \bm r^\top)^\top  \rangle ,
\end{align}
in terms of the quadrature vector $\bm{r}$.
From Eqs.~\eqref{eq:S-matrix} and \eqref{eq:K-matrix} we can 
define a linear transformation between input and output quadrature operators
\begin{align}
\bm{r}_{\rm out} = A \bm{r}_{\rm in} ,
\end{align}
from which we have
\begin{align}
\bm r_{\rm out} \cdot \bm r_{\rm out}^\top =  A \bm{r}_{\rm in} (A \bm{r}_{\rm in})^\top = A (\bm r_{\rm in} \cdot \bm r_{\rm in}^\top) A^\top . \nonumber
\end{align}
Therefore,  the input and output covariance matrices relate as
\begin{align}
V_{\rm out} = A V_{\rm in} A^\top. \label{eq:v_out_theory}
\end{align}
If we assume the input modes to be in the vacuum state, then $V_{\rm in}  = \mathbb{I}$.

The theory matrix in Fig.~3 is given by Eq.~\eqref{eq:v_out_theory}, where we assume the input state to be vacuum.
Hence, the output covariance matrix $V_{\rm out}$ is determined by matrix $A$, which is in turn set by the parameters $\mu_1$, $\mu_2$, $\omega_n$, $\kappa$ and $\tilde{\omega}_r$.
Since the parameters $\kappa = 2\pi \cdot \SI{124}{\mega\hertz}$ and $\tilde{\omega}_r = 2\pi \cdot \SI{4.3}{\giga\hertz}$ are obtained from a frequency spectroscopy sweep, and $\omega_n$ is set by the experimentalist, this leaves us with only two parameters, $\mu_1$ and $\mu_2$, which are complex effective couplings between modes and controlled by the two flux pumps.
We adjust these values by hand, obtaining good correspondence between experiment and theory with $\mu_1 = 2\pi \cdot 15.6\angle90^{\SIUnitSymbolDegree}\;\si{\mega\hertz}$ and $\mu_2 = 2\pi \cdot 15.6\angle-60^{\SIUnitSymbolDegree}\;\si{\mega\hertz}$.

\subsection{Gaussianity of measured noise}

\begin{figure*}
    \centering
    \includegraphics[scale=0.7]{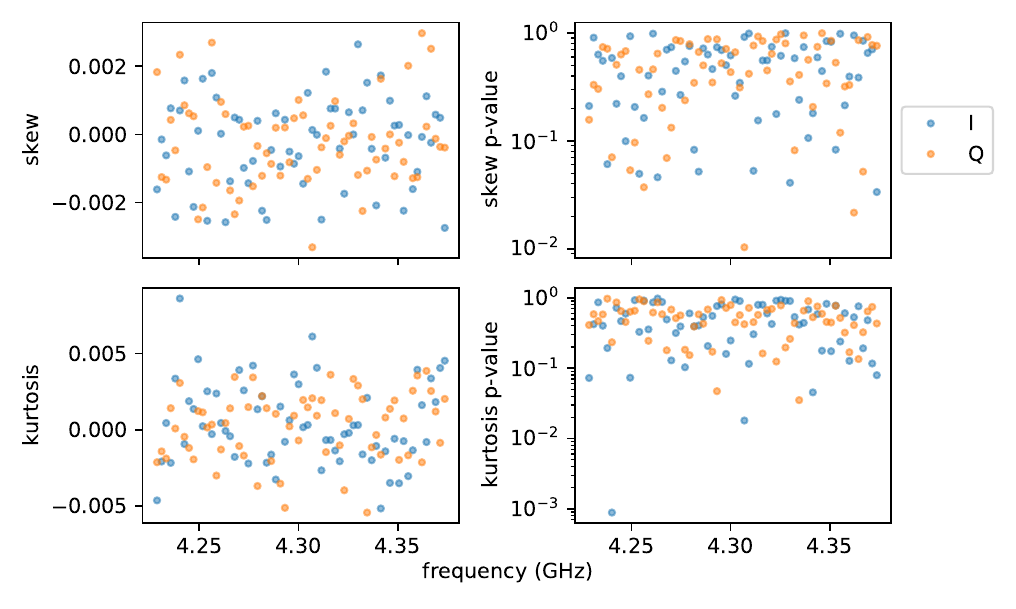}
    \caption{
    Skew and kurtosis of our sampled distributions, and the corresponding p-values with our null hypothesis being that our underlying distribution is Gaussian. All skew and kurtosis values are small (ideally zero), and the corresponding p-values mostly greater than $0.1$. Hence we deduce that distribution underlying our measured data is indeed Gaussian.
    }
    \label{fig:gauss_test}
\end{figure*}

Our theoretical model and experimental analysis considers the noise to be Gaussian.
A distribution is Gaussian if it is completely characterized by the first two moments.
We therefore check for Gaussianity by also calculating the third and fourth order moments in our data, called the skewness and kurtosis.  We use the \texttt{scipy.stats}-package \cite{scipy_skewness, scipy_kurtosis} to test the distribution of $I$- and $Q$-quadratures at each frequency.
The result is presented in Fig.~\ref{fig:gauss_test}.
Skew and kurtosis are zero for an ideal Gaussian distribution.
We see that all values (for all quadratures and all frequencies) are small.
But are these values are small enough given the number of data points?
To answer this question we do a skew and kurtosis test, where our null hypothesis is that our data is drawn from a Gaussian distribution.
The right-hand column in Fig.~\ref{fig:gauss_test} presents the p-value for this hypothesis test.
Since a majority of the p-values are larger than $0.1$ we have a high level of confidence that our data is sampled from a Gaussian distribution.

\section{Multipartite Entanglement}

\subsection{Bipartite entanglement criteria}

In Fig.~2b we compared three entanglement tests: the Duan criterion, the PPT test and the SvL test.
Consider two Gaussian states $A$ and $B$ residing in the joint Hilbert space $\mathcal{H}_A \otimes \mathcal{H}_B$.
With our choice of normalization, the Duan criterion is \cite{Duan2000}
\begin{align}
    \langle (\Delta u)^2 \rangle + \langle (\Delta v)^2 \rangle - 4 \geq 0,
\end{align}
where $u = x_A + x_B$ and $v = p_A - p_B$. 
The quadrature operators $x_{A/B}$ and $p_{A/B}$ satisfy the commutation relations $[x_A, p_A]=[x_B, p_B]=2i$ and the  mean is taken over the bipartite state described by the covariance matrix $V_{AB}$.

The Duan criteria is a necessary and sufficient condition for separability.
Another necessary and sufficient condition for separability is the Positive Partial Transpose (PPT) test \cite{Simon2000}
\begin{align}
    V^{\rm PPT} + i\Omega \geq 0,
\end{align}
where $V^{\rm PPT} = \Lambda V_{AB} \Lambda$ and $\Lambda = \rm{diag}(1,1,1,-1)$.
Both Duan and PPT criteria apply to bipartite systems and they can not be be used to determine multipartite entanglement.  For multipartite entanglement we use the SvL criteria, which tests for arbitrary $k$-partite entanglement. 

\subsection{Genuine multipartite entanglement (GME) and full inseparability}

An $n$-partite mixed state $\rho$ is called \emph{fully separable} if it can be written as a convex combination of product states:
\begin{equation}
    \rho = \sum_i p_i \left(\rho_1^i \otimes \rho_2^i \otimes \cdots \otimes \rho_n^i\right),
\end{equation}
where $p_i \ge 0$ and $\sum_i p_i =1$.
This is a direct generalization of the definition of separability for bipartite states $n = 2$. 
In contrast to a bipartite state, which is either entangled or separable, there exists several classes of entanglement and separability for multipartite states.
One can arrange the $n$ parts (modes) into $k\le n$ partitions, which are then considered as subsystems.
States that are fully separable with respect to this partition (or can be written as a mixture of such states) are called \emph{$k$-separable}. A state that is not $k$-separable is called $k$-\emph{inseparable}~\cite{Seevinck2008Sep}.
A state that is not separable with respect to any such split is called fully ($n$-partite) \emph{inseparable}~\cite{Dur2000Mar}.

Let us consider a tripartite state as an example. It is fully separable if it can be decomposed as
\begin{equation}\label{eq:tripartite_fullysep}
    \rho = \sum_i p_i \left(\rho_1^i \otimes \rho_2^i \otimes \rho_3^i\right).
\end{equation}
It is 2-separable or \emph{biseparable} if it can be written as either of the following states, or a mixture thereof:
\begin{align}\label{eq:tripartite_bisep}
        \rho &= \sum_i p_i\left( \rho_{12}^i \otimes \rho_3^i\right),\nonumber\\
        \rho &= \sum_i p_i\left( \rho_{13}^i \otimes \rho_2^i\right),\\
            \rho &= \sum_i p_i\left( \rho_{23}^i \otimes \rho_1^i\right),\nonumber
            \end{align}
where $\rho_{12}^i$, $\rho_{13}^i$ and $\rho_{23}^i$ are bipartite states. They can be either entangled or separable but in the latter case we will actually have the above fully separable case. In general $(k+1)$-separability implies $k$-separability~\cite{Hong2016Apr}.
Note that if the state is a statistical mixture of all three states~\eqref{eq:tripartite_bisep}, there is no \emph{single} bipartite split with respect to which the state is separable, but the state is nonetheless biseparable since it is a mixture of biseparable states~\cite{Huber2010May}.

If a state is not biseparable, it is said to be \textit{genuinely multipartite entangled} (GME). That is because it implies entanglement for any other partitioning of modes. For pure states, GME is equivalent to full inseparability. This is not the case for mixed states. In this context, GME implies full inseparability, but not the other way around. Hence GME is a stronger form of correlation \cite{Teh2019Aug}.

The SvL test result of Fig.~4 show that none of our bipartitions are separable, which suggests we are generating fully inseparable states \cite{Gerke2016, Shalm2013Jan, Teh2014Dec}, but not necessarily GME as we have not ruled out mixtures of biseparable states. 

\subsection{GME in lossy systems}

\begin{figure}[htbp]
    \centering
    \includegraphics[width=0.9\linewidth]{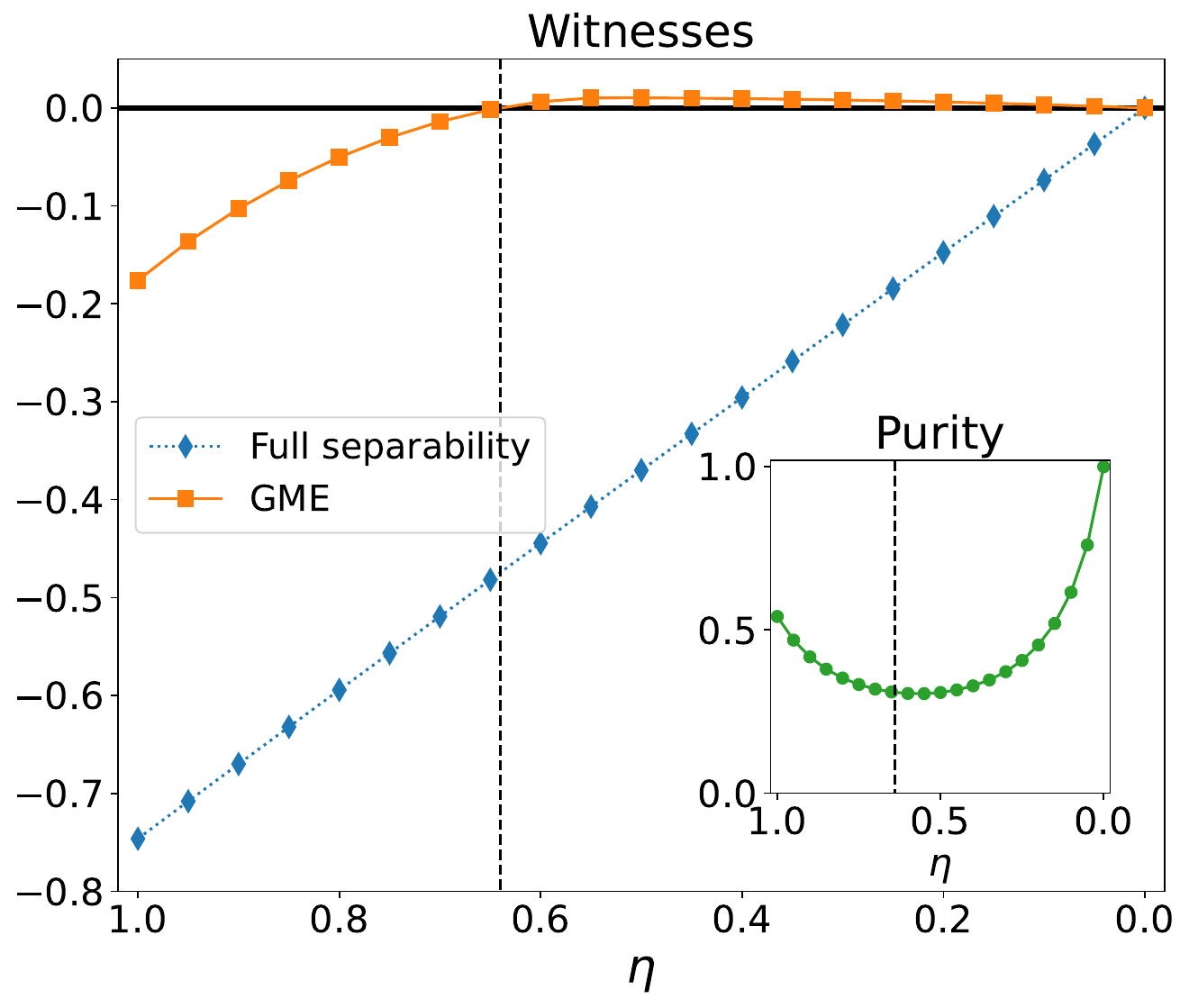}
    \caption{Witnesses of full separability and GME for increasing amounts of vacuum mixed into an ideal state of five even modes, as a function of beam splitter transmittivity $\eta$. The state is never fully separable except for $\eta=0$ when the state is completely vacuum, but GME is only present until the purity has decreased to $0.31$, indicated by the vertical dashed line. The inset shows purity which decreases as more vacuum is mixed into the state, and then increases after $\eta=0.5$ as the state becomes more and more dominated by vacuum.}
    \label{fig:purity_GME}
\end{figure}

It is known that testing for GME is more sensitive to imperfections of the experiment, in comparison with other measurements of entanglement~\cite{Guhne2009Apr, Aoki2003Aug}.
To explore this effect we simulate losses by mixing a signal with an auxiliary mode in the vacuum or a thermal state at a fictitious beam-splitter. After the beam-splitter, the reflected mode is lost and we are left with the transmitted one. The beam-splitter transmissivity $\eta$ determines what fraction of the original signal is transmitted with $\eta = 1$ corresponding to the lossless case while $\eta < 1$ implies the presence of losses, which in an experimental setting could stem from cryogenic electronics such as cables and isolators.
We combine $n$ modes from an ideal covariance matrix $V_{\rm{ideal}}$ (calculated according to the theory in Sec.~\ref{sec: cov matrix theory}) with vacuum $\mathbb{I}_{2n}$. 
The covariance matrix of the beam-splitter input is the direct sum of $V_\textbf{ideal}$ and $I_{2n}$
\begin{equation}
V=
    \begin{pmatrix}
        V_\textbf{ideal} & 0 \\
        0 & \mathbb{I}_{2n}
    \end{pmatrix}.
\end{equation}
The state transforms as
\begin{equation}
    V' = SVS^\top
\end{equation}
under the beam-splitter transformation $S$, which is given by~\cite{Serafini2017Oct}
\begin{equation}
\begin{pmatrix}
    \cos\theta \mathbb{I}_{2n} & \sin\theta\mathbb{I}_{2n} \\
    -\sin\theta \mathbb{I}_{2n} & \cos\theta \mathbb{I}_{2n}
\end{pmatrix},
\end{equation}
with $\eta=\cos^2\theta$. 
After the vacuum modes are mixed in and traced out, we select a subset of 5 modes and calculate the purity $p(V)=1/\sqrt{|V|}$ \cite{Serafini2017Oct} as well as the entanglement witness. Here we use the program \texttt{FullyWit} that tests for full separability, and \texttt{MultiWit} that tests for GME, both from Ref.~\cite{Hyllus2006Apr}.  The results shown in Fig.~\ref{fig:purity_GME} for different levels of vacuum noise.  

Selecting only a subset of modes from a larger entangled system causes a reduction of purity. In this example with five modes, the initial purity is $p=0.5410$. In Fig.~\ref{fig:purity_GME}, we show how the witnesses of full separability and GME change as $\eta$ is reduced, meaning more vacuum is mixed into the state. When the full separability witness is negative, it means the state is not fully separable, i.e.\ there is some type of entanglement in the state. If the GME witness is negative it means the state is genuinely multipartite entangled. It can be seen that the full separability witness increases linearly with the reduced $\eta$, but remains negative until $\eta=0$ when the state is fully vacuum. However, the state is genuinely multipartite entangled only until $\eta=0.64$ corresponding to a purity of $0.31$.

Our example shows that it is in theory possible to generate GME states using our method, provided that the purity is not too low.
An analysis of our data however, reveals that the $K^{-1}$ sets have an average purity of $0.20$, while for $K^{1}$ and $K^{0}$ it is $0.16$ and $0.06$ respectively.
It is therefore unlikely our current dataset exhibits GME.
An application of the Hyllus and Eisert test~\cite{Hyllus2006Apr} to our data provides no strong evidence for GME.
Future research should improve on the experimental methodology to enable states with higher purity.

\bibliographystyle{ieeetr}
\bibliography{refs_supplementary}